\documentclass[preprint,12pt]{elsarticle}

\usepackage{amssymb}
\usepackage{amsmath}
\usepackage{hyperref}
\usepackage{float}
\usepackage[caption=false]{subfig}%
\usepackage{algorithm}%
\usepackage{algorithmicx}%
\usepackage{algpseudocode}%

\journal{Applied Soft Computing}

\begin{document}

\begin{frontmatter}

\title{Multi-agent Auditory Scene Analysis}

\author[1]{Caleb Rascon} %% Author name
\ead{caleb@unam.mx}
\affiliation[1]{organization={Instituto de Investigaciones en Matematicas Aplicadas y en Sistemas, Universidad Nacional Autonoma de Mexico},%Department and Organization
            addressline={Circuito Escolar 3000}, 
            city={Coyoacan},
            postcode={04510}, 
            state={Ciudad de Mexico},
            country={Mexico}}

\author[2]{Luis Gato-Diaz} %% Author name
\ead{luisgatodiaz@microsoft.com}
\affiliation[2]{organization={M365 Core, Microsoft Corporation},%Department and Organization
	addressline={1 Microsoft Way}, 
	city={Redmond},
	postcode={98052}, 
	state={Washington},
	country={United States}}

\author[3]{Eduardo Garc\'{i}a-Alarc\'{o}n} %% Author name
\ead{egarciaa@comunidad.unam.mx}
\affiliation[3]{organization={Posgrado en Ciencias e Ingenieria de la Computacion, Universidad Nacional Autonoma de Mexico},%Department and Organization
	addressline={Circuito Escolar 3000}, 
	city={Coyoacan},
	postcode={04510}, 
	state={Ciudad de Mexico},
	country={Mexico}}

%% Abstract
\begin{abstract}
Auditory scene analysis (ASA) aims to retrieve information from the acoustic environment, by carrying out three main tasks: sound source location, separation, and classification. These tasks are traditionally executed with a linear data flow, where the sound sources are first located; then, using their location, each source is separated into its own audio stream; from each of which, information is extracted that is relevant to the application scenario (audio event detection, speaker identification, emotion classification, etc.). However, running these tasks linearly increases the overall response time, while making the last tasks (separation and classification) highly sensitive to errors of the first task (location). A considerable amount of effort and computational complexity has been employed in the state-of-the-art to develop techniques that are the least error-prone possible. However, doing so gives rise to an ASA system that is non-viable in many applications that require a small computational footprint and a low response time, such as bioacoustics, hearing-aid design, search and rescue, human-robot interaction, etc. To this effect, in this work, a multi-agent approach is proposed to carry out ASA where the tasks are run in parallel, with feedback loops between them to compensate for local errors, such as: using the quality of the separation output to correct the location error; and using the classification result to reduce the localization's sensitivity towards interferences. The result is a multi-agent auditory scene analysis (MASA) system that is robust against local errors, without a considerable increase in complexity, and with a low response time. The complete proposed MASA system is provided as a publicly available framework that uses open-source tools for sound acquisition and reproduction (JACK) and inter-agent communication (ROS2), allowing users to add their own agents.\end{abstract}

%\includegraphics[width=\textwidth]{GraphicalAbstract}

%\item The linear data flow of typical auditory scene analysis frameworks are sensitive to local errors, requiring high complexity for each its task techniques, producing high latency and large computational footprint.
%\item A multi-agent approach is proposed with feedback loops between tasks, providing robustness against local errors, requiring less complex task techniques, providing lower latency and less computational requirements.
%\item The proposed approach is adaptive by nature and is publicly available at \url{https://github.com/balkce/masa}, using open-source tools for sound acquisition and inter-task communication.

%% Keywords

\end{frontmatter}

%\linenumbers


\section{Introduction}\label{sec:intro}%% Labels are used to cross-reference an item using \ref command.

Auditory scene analysis (ASA) aims to provide a description of the sound sources in the environment \cite{bregman1994auditory}, including their location, their separated audio stream, and their type (human, urban sound, noise, etc.). These tasks are of interest to be emulated computationally \cite{wang2006computational}, which can benefit several application scenarios such as bioacoustics \cite{stowell2017computational}, hearing aid research \cite{zhang2019study,green2022speech}, search and rescue scenarios \cite{nakadai2017development}, human-robot interaction \cite{rascon2015integration}, etc.

Usually, ASA is divided into three main tasks: to locate the sound source in the environment (localization) \cite{rascon2017localization}; to separate the audio data from each sound source in their own stream (separation) \cite{das2021fundamentals}; and to extract information from each sound source that describes it (classification), such as the type of sound source \cite{martin1999sound,salamon2015unsupervised}, who is speaking \cite{campbell2002speaker}, what is that person saying \cite{eskimez2018front}, their emotional state \cite{basu2017review}, etc.

\begin{figure}[H]
	\centering
	\includegraphics[width=0.65\textwidth]{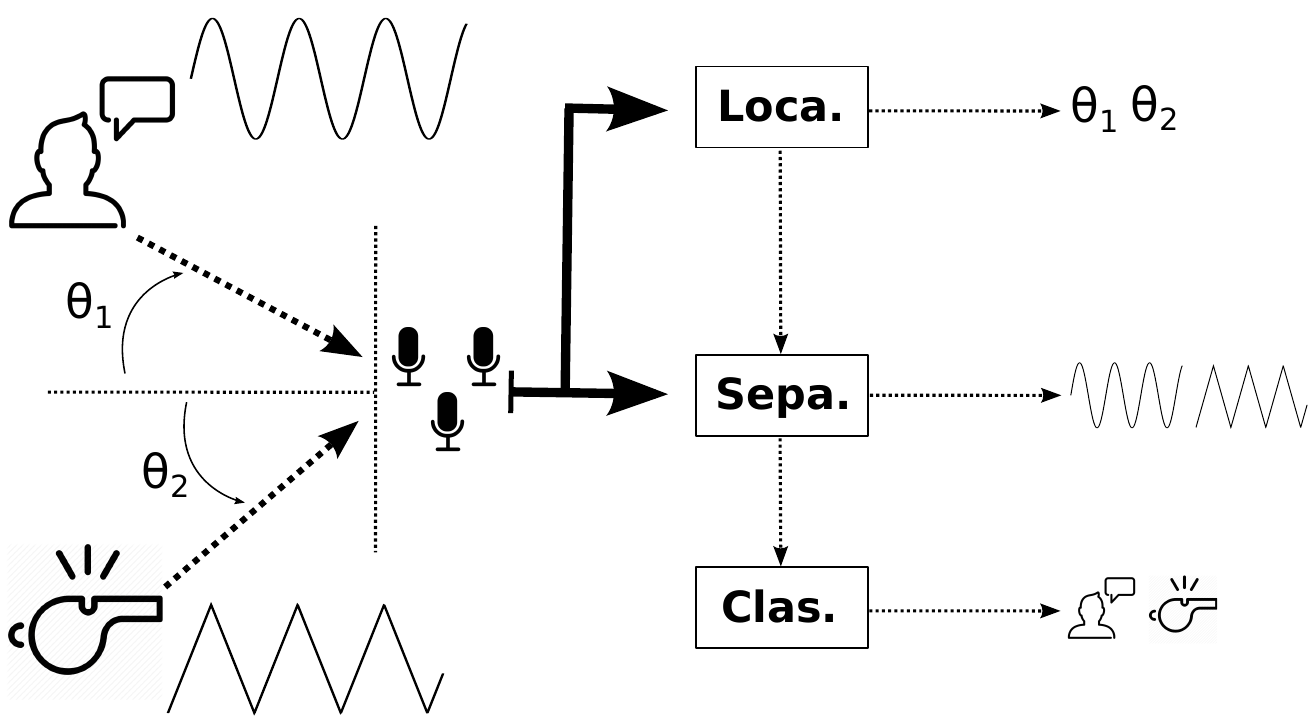}
	\caption{Typical data flow for auditory scene analysis.}\label{fig:typicalasa}
\end{figure}

Current ASA frameworks, such as HARK \cite{nakadai2017hark} and ODAS \cite{grondin2022odas}, typically employ a data flow between tasks that is linear in nature, as exemplified in Figure \ref{fig:typicalasa}.

There are several limitations that come with this type of data flow. First off, if an error occurs at the beginning of the data flow, such an error is `passed on' to the following tasks. Meaning, a localization error begets separation degradation, which in turn begets a wrong classification. To tackle this issue, local errors (those that occur at the task level) are usually minimized by making the tasks' techniques robust against the local errors of prior tasks. This has the added issue that complexity is added onto the tasks' techniques, which requires increasing the use of computational resources. Additionally, this added complexity may increase the technique's response time, which is further aggravated by the fact that, in linear data flows, the overall response time is the sum of the response times of all the tasks. An increase to this overall response time may limit the system's viability in real-time\footnote{For simplicity, the terms ``real-time'' and ``online processing'' are used inter-changeably in this work.} applications  \cite{leinbaugh1980guaranteed,joseph1986finding}. And, finally, in linear data flows, the adaptability of the overall systems relies on the robustness against change of the tasks' techniques: if one of these isn't robust, the whole system isn't.

To this effect, in this work, a new paradigm to carry out ASA is proposed, where it is structured as a multi-agent system (MAS) \cite{dorri2018multi}. This type of systems aim to model very complex behaviors as a set of small computing entities (agents) that run in parallel, and that solve smaller, simpler tasks. The key to a MAS is that these agents are expected to interact with one another \cite{van2008multi}, employing a non-linear information data flow. This inter-agent interaction is from which the complex behavior is expected to emerge \cite{dorri2018multi}. Using a MAS to solve a complex task has several benefits, such as: efficient operations due to its parallelization potential, reliability and robustness, agents can be added or removed to fit the needs of the application scenario, less computational cost compared to a centralized approach, etc. \cite{balaji2010introduction}.

\begin{figure}[H]
	\centering
	\includegraphics[width=0.65\textwidth]{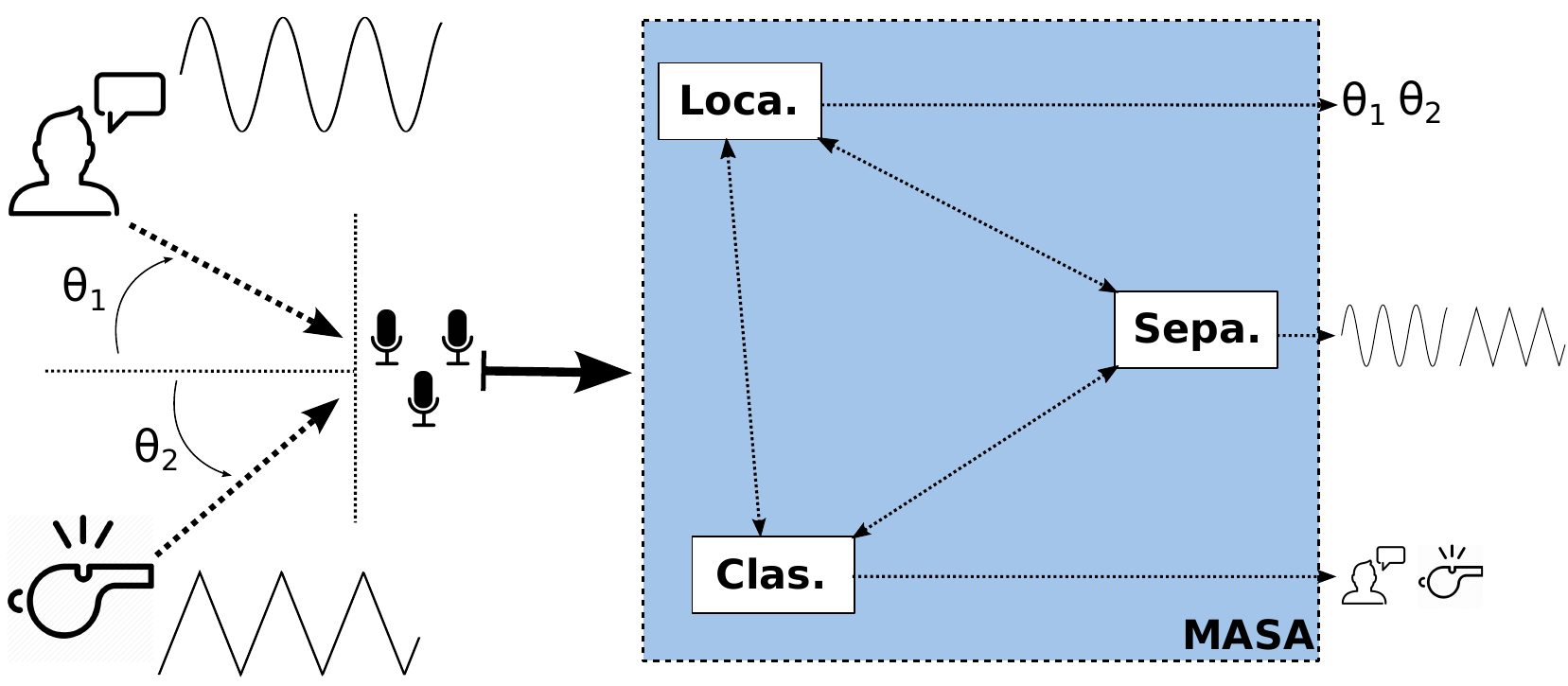}
	\caption{General diagram of the proposed multi-agent auditory scene analysis (MASA) approach.}\label{fig:masa}
\end{figure}

For the case of ASA, in Figure \ref{fig:masa}, a general proposal of a multi-agent approach (MASA) is shown.

The specific benefits from structuring ASA as a MAS are as follows:

\begin{itemize}
	\item The feedback nature of a MAS removes the necessity of robustness in the technique itself, since robustness is ``taken care of'' at the system level.
	\item Although there is complexity added at the local level to consider inter-agent information inside the agent's technique, it is far lesser than the complexity required to make it robust against local errors of other agents.
	\item The overall response time becomes that of the slowest agent, which is considerably lower than that of a linear data flow (the sum of all agent's response times).
	\item The system is built from the ground up to be adaptive.
\end{itemize}

Recently, an effort into this multi-agent paradigm was carried out \cite{rascon2025direction} where a feedback loop was implemented between the separation and the localization tasks, to correct location errors in real-time based on speech quality. It showed strong robustness against location errors of up to $20^\circ$, which is strong evidence of the viability of the proposed multi-agent approach. In the work presented here, the effort of \cite{rascon2025direction} is expanded upon to formalize it into a complete framework, with several local improvements to its currently implemented agents and additional feedback inter-agent loops. Specifically, the contributions of this work are as follows:

\begin{itemize}
	\item A generalized multi-agent framework for auditory scene analysis (MASA).
	\item A feedback loop between the classification and localization tasks, to make the system robust against specific types of urban-sound interferences.
	\item Improvements on the techniques used in several of MASA's agents, compared to their previously published versions.
	\item Given these improvements, and the new feedback loop, location errors are reduced (compared to \cite{rascon2025direction}), as well as a measured increase in speech quality.
\end{itemize}

The proposed framework was built using open source available protocols for audio acquisition and reproduction (JACK \cite{JACKAudio}) and inter-agent communication (ROS2 \cite{ROS2}), and its full implementation is publicly available at \url{https://github.com/balkce/masa}. This provides users the possibility to add their own agents to the MASA framework.

The paper is structured as follows: Section \ref{sec:proposed} presents the proposed multi-agent approach, along with the description of each of MASA's agents and new local improvements; Section \ref{sec:implementation} describes some implementation aspects, such as the tools necessary to run MASA, as well as additional quality-of-life agents; Section \ref{sec:results} compares the overall performance of running ASA with a traditional linear data flow, with one classification-to-location feedback loop, and the full proposed MASA approach; and Section \ref{sec:conclusions} presents some concluding remarks, as well as some issues that are left to be tackled as future work.

\section{Proposed Multi-agent Approach}\label{sec:proposed}

\begin{figure}[H]
	\centering
	\includegraphics[width=0.65\textwidth]{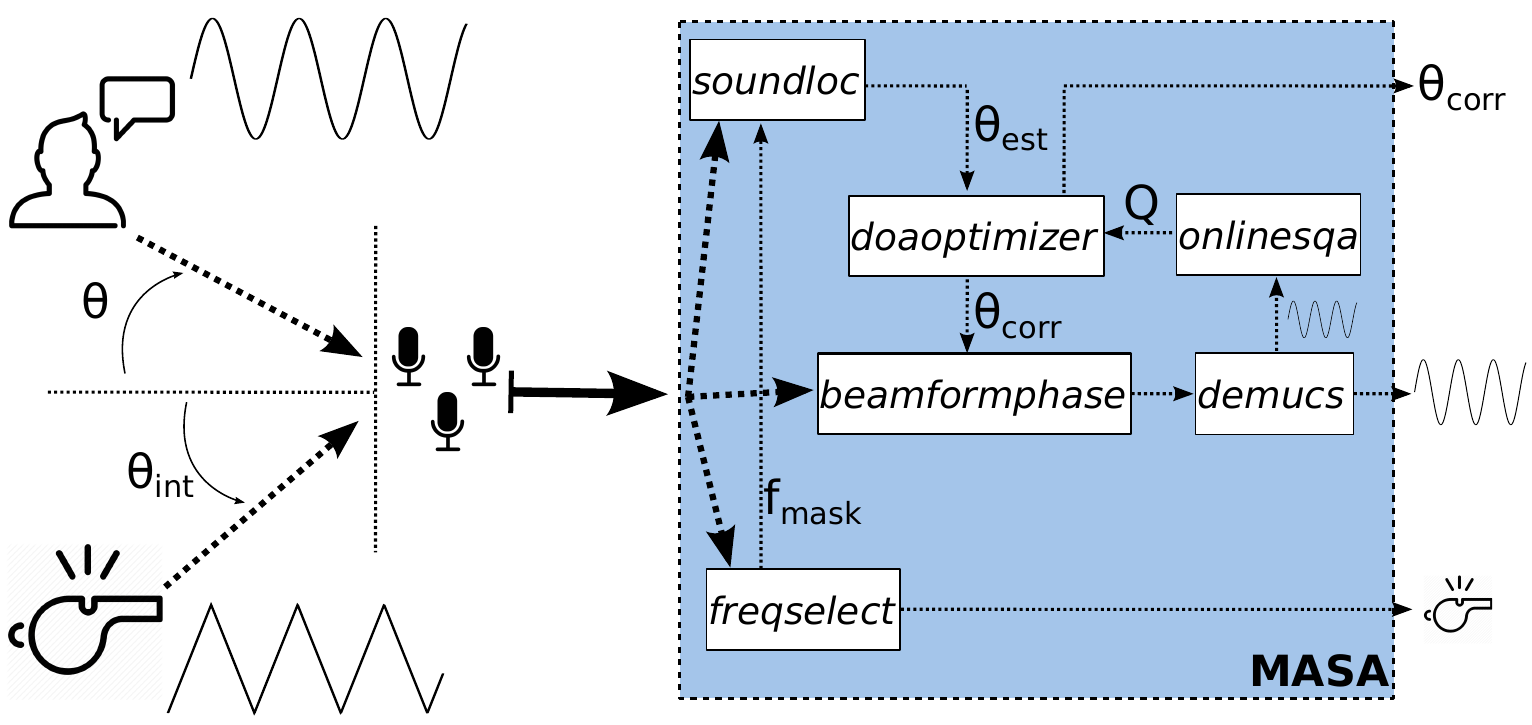}
	\caption{Currently implemented version of multi-agent auditory scene analysis (MASA).}\label{fig:currentmasa}
\end{figure}

In Figure \ref{fig:currentmasa}, the version of the proposed multi-agent auditory scene analysis (MASA) that is currently implemented is presented.

All the agents can be seen that are currently a part of MASA, as well as how they are inter-connected. Each of these agents is detailed in the following sections.

\subsection{Sound Source Localization}

\textbf{Agent name:} \textit{soundloc}.

This agent is based on the work of \cite{rascon2015lightweight}, where a lightweight technique was proposed to estimate the direction of arrival (DOA) of multiple mobile speech sources using only three microphones. It was shown to be able to estimate the DOA of up to 4 simultaneous speech sources, and was able to track several mobile sound sources. It assumes that simultaneous sources do not fully overlap, such that several DOA ``candidates'' are gathered over time. However, the simple manner in which these candidates are estimated, although effective, left a performance gap in need of closing.

Building on \cite{rascon2015lightweight}, a novel multi-speaker localization and tracking technique was developed using a triangular microphone array \cite{gato2020localizacion}, which is described here for completeness sake. In this technique, for each microphone pair:

\begin{enumerate}
	\item The Fast Fourier Transform (FFT) is applied to each input time window ($1024$ samples at $48$ kHz $\sim 21.3$ ms).
	\item A band-pass filter is applied to each input channel, to remove information from frequencies outside the 1–4 kHz range (targeting human speech).
	\item The generalized cross-correlation with phase transform (GCC-PHAT) \cite{knapp2003generalized} is applied to each input channel pair.
	\item Correlation peaks are identified and mapped to possible azimuth DOAs.
	\item Each DOA is transformed into a common reference frame.
	\item A coherence metric is calculated by the average difference between all possible DOA combinations.
	\item A DOA combination is deemed coherent (below the $\pm$15$^\circ$ coherence threshold), and used to determine the direction of the speaker.
\end{enumerate}

GCC-PHAT enhances robustness against reverberation, while frequency filtering reduces noise-induced false positives. This also enables multi-speaker detection via clustering of arrival directions, since only coherent DOAs are provided to the tracking side of the localization technique. More specifically, tracking goes as follows:

\begin{enumerate}
	\item A historical record of coherent DOAs is maintained.
	\item K-means clustering \cite{jin2017k} is used to group DOAs into clusters, each representing one speaker, effectively carrying out multiple-DOA estimation through time.
	\item A Kalman filter \cite{kalman1960filter} is applied to each DOA cluster.
	\item A circular accelerated motion model is employed for prediction.
	\item Updates are then estimated with new localization data.
\end{enumerate}

Speaker movement is modeled as circular motion, with variable velocity and normally distributed acceleration. This provides better adaptability than constant-velocity models, while being more efficient than complex non-linear alternatives. Additionally, the size of each DOA cluster serves as a type of confidence value of each DOA estimation.

This localization system was compared with another popular lightweight sound source localization system, which is part of the ODAS framework \cite{odas2019}. It employs beamforming for source localization instead of TDOA estimation, as well as particle filtering for source tracking instead of Kalman filtering. This comparison employed the AIRA corpus \cite{rascon2018acoustic}, with recordings being fed to both systems in real-time (to emulate a real-life setting), in four different scenarios (an anechoic chamber, a student cafeteria, a retail store, and an office corridor) with varying reverberation times (10–270 ms), signal-to-noise ratios (10–43 dB), number of speakers (1 to 4), type of speakers (stationary and moving), and type of microphone array (a 3-microphone triangular array, same as \cite{gato2020localizacion}; and an 8-microphone circular array, same as \cite{odas2019}).

The performance was measured in terms of number of speakers detected and DOA accuracy. The computational efficiency of both systems was also registered, and measured as the CPU usage percentage of one Intel Core i5-7200U core. The results showed that in reverberant/noisy scenarios, the performance gap between the proposed system and ODAS using 8 microphones was comparable. Given that ODAS was later improved in \cite{odas2022}, this performance gap has potentially narrowed even further. However, ODAS with 3 microphones performed worse in detection and false positive rate when tested under identical conditions. Additionally, there was a $\sim 50$\% reduction in computational requirements, since \cite{gato2020localizacion} only used $\sim 35$\% of the CPU with 3 mics, while \cite{odas2022} used $\sim 70$\%. Given its relatively high performance, with its low computational requirements, the technique presented in \cite{gato2020localizacion} is the one being used in the \textit{soundloc} agent.

\subsection{Speech Enhancement}
\textbf{Agents names:} \textit{beamformphasemix} and \textit{demucsmix}.

\begin{figure}[H]
	\centering
	\includegraphics[width=0.7\textwidth]{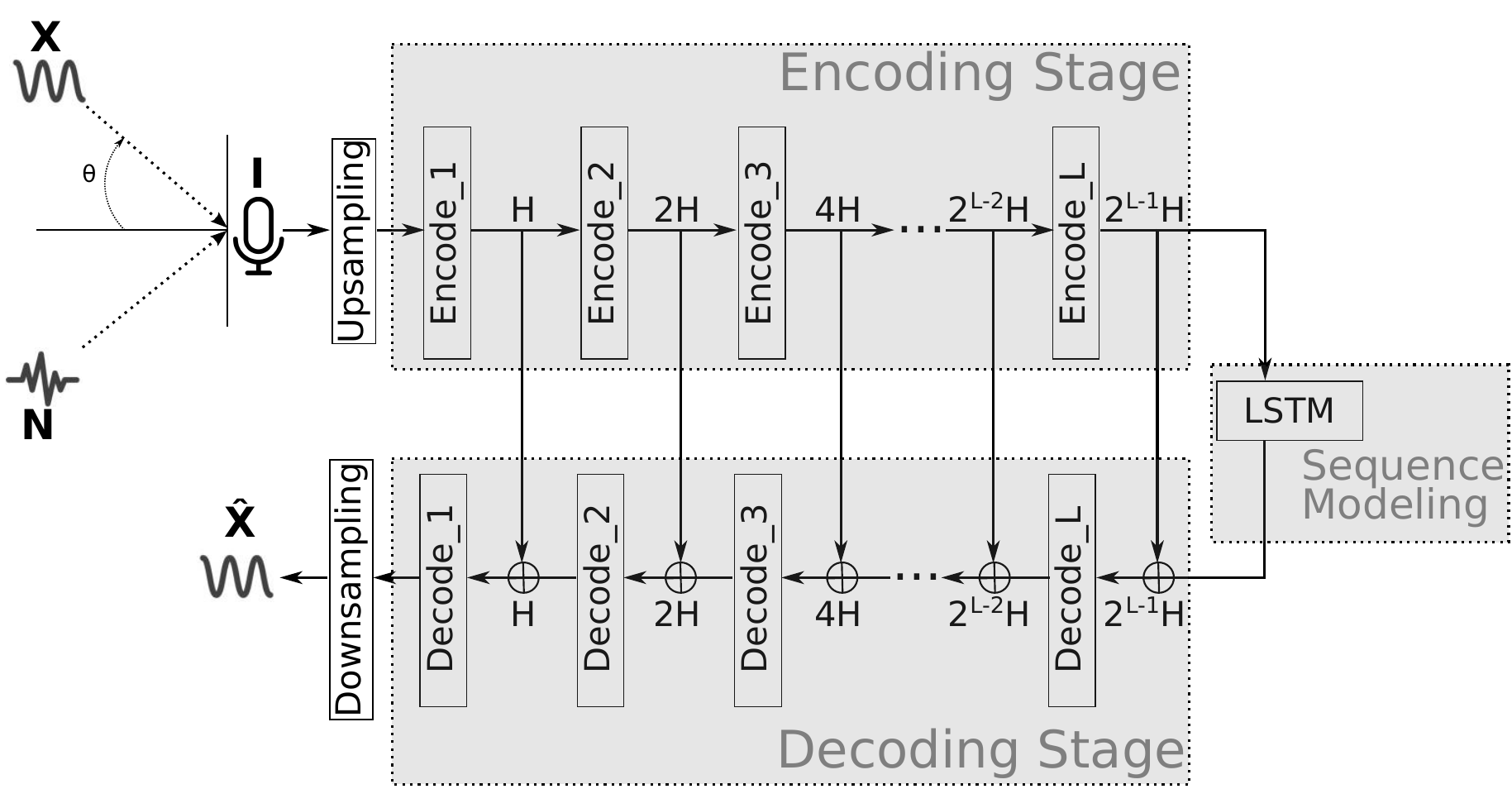}
	\caption{Architecture of the Demucs denoiser model.}\label{fig:demucs}
\end{figure}

\begin{figure}[H]
	\centering
	\subfloat[Encode module.]{\includegraphics[width=0.4\textwidth]{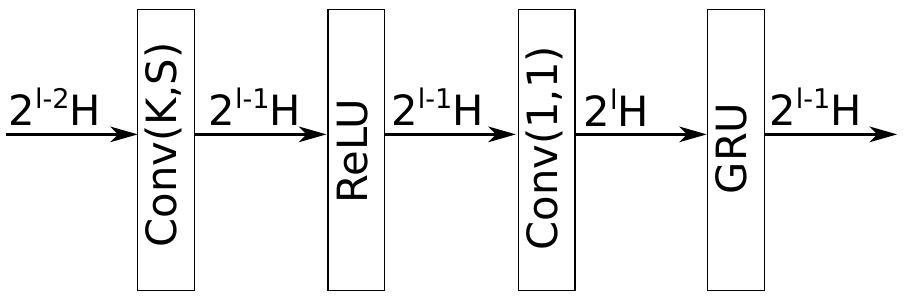}\label{fig:encode}}
	\hspace{1cm}
	\subfloat[Decode module.]{\includegraphics[width=0.4\textwidth]{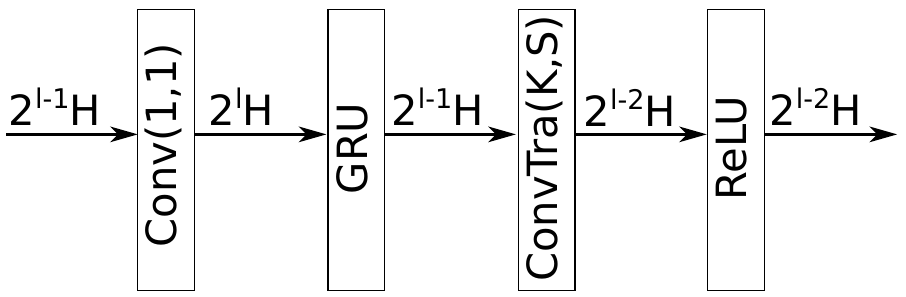}\label{fig:decode}}
	\caption{Architectures of the encode and decode modules.}\label{fig:decencmods}
\end{figure}

Real-time speech enhancement has improved recently, in great part because of lightweight models such as Demucs \cite{defossez20_interspeech}, which has an architecture based on the U-Net paradigm \cite{ronneberger2015u}, as shown in Figure \ref{fig:demucs}. In Figures \ref{fig:encode} and \ref{fig:decode}, the architectures of the encoding and decoding models are shown.

\begin{figure}[H]
	\centering
	\includegraphics[width=0.7\textwidth]{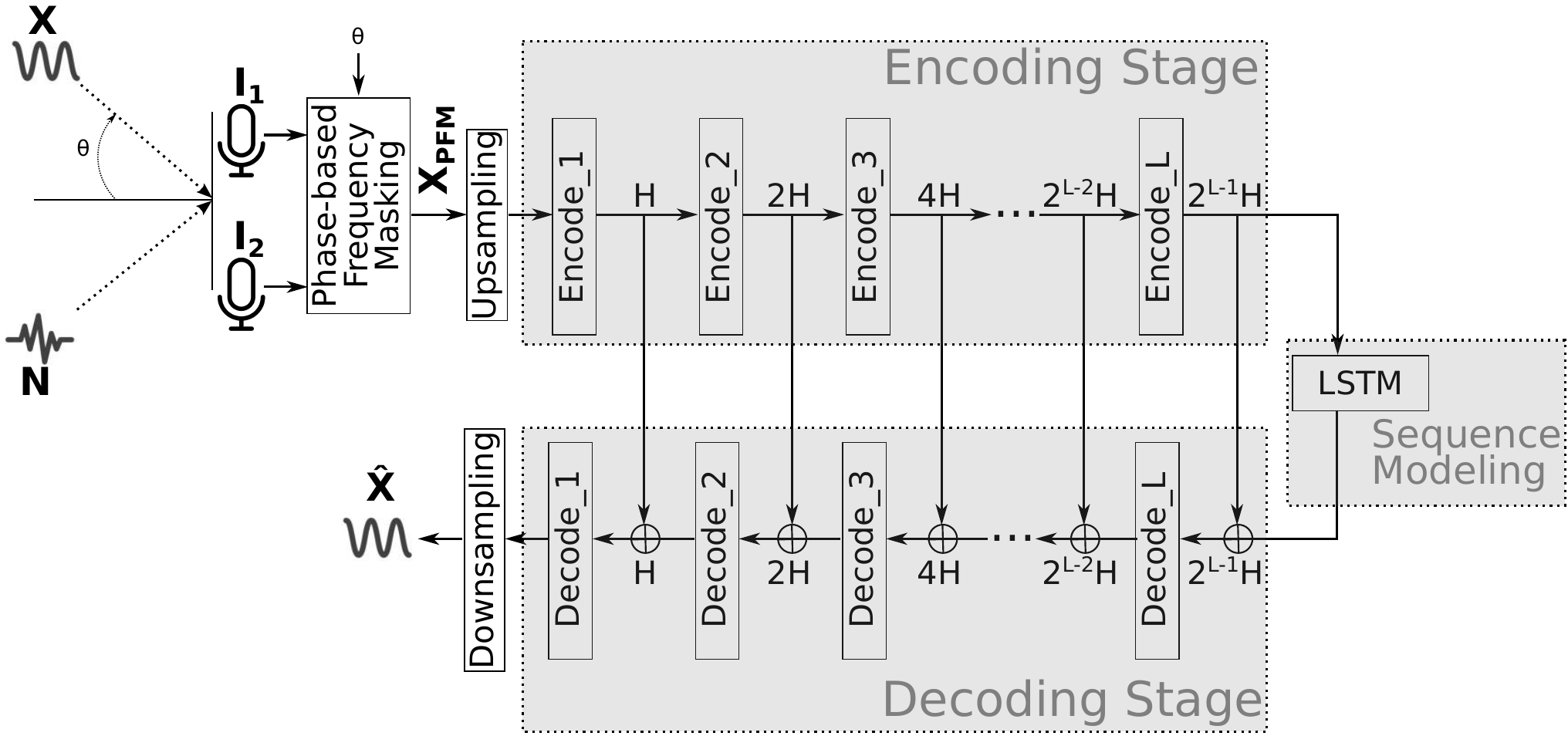}
	\caption{Diagram of the location-based target selection strategy \cite{rascon2023target}.}\label{fig:demucslocate}
\end{figure}

However, it has been shown that it performance drops substantially in multi-speaker scenarios \cite{rascon2023characterization}. This is understandable, given the ``one speech source to enhance'' assumption that most (if not all) speech enhancers are trained with. However, a phase-based frequency-masking beamformer \cite{rascon2021corpus} can be used to ``nudge'' the Demucs model towards the speech source of interest. The beamforming output increases the energy of the steered source (although it does not separates it completely from the rest of the noisy mixture). In conjunction, it has been shown that the Demucs model tends to separate the highest-energy speech source \cite{rascon2023characterization} in multi-speaker scenarios. Thus, a real-time location-based target selection strategy has been previously proposed \cite{rascon2023target} which steers the Demucs model (here referred as \textit{demucs}) to enhance a speech source that is located at a given DOA. The complete location-based target selection strategy described in \cite{rascon2023target} is summarized in Figure \ref{fig:demucslocate}.

Unfortunately, this strategy has been shown to be sensitive to location errors \cite{rascon2023target}. This issue prompted a real-time correction system of the DOA by maximizing the quality of the enhanced speech \cite{rascon2025direction}, which is the predecessor of the multi-agent approach here proposed. However, the enhancement model can still be improved upon locally.

\begin{figure}[H]
	\centering
	\includegraphics[width=0.69\textwidth]{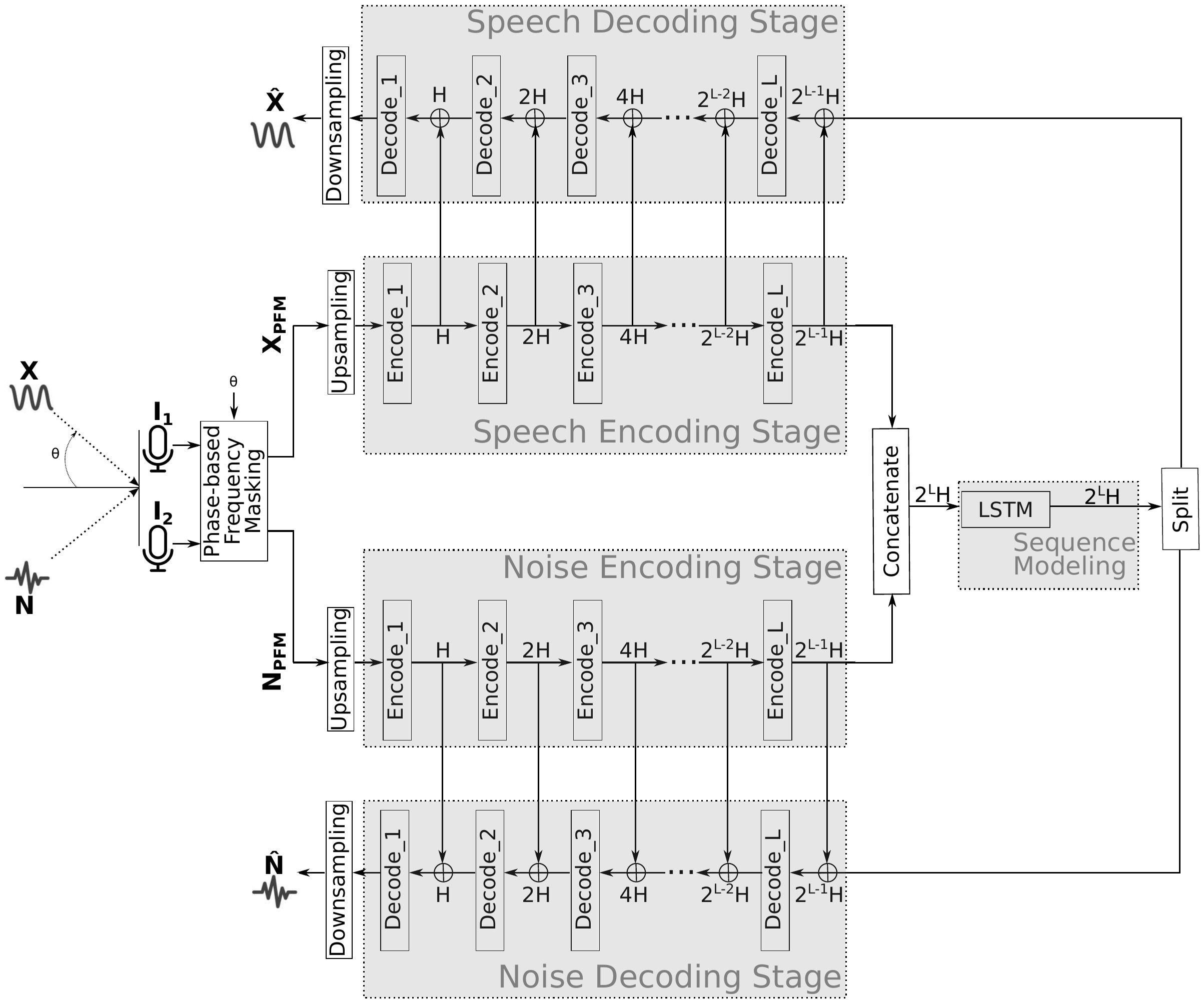}
	\caption{Architecture of the Demucs denoiser model that uses both the preliminary estimation of the target speech, as well as of the cumulative environmental interference.}\label{fig:demucslocatemix}
\end{figure}

In the work presented here, the \textit{demucs} speech enhancement model is modified so that it takes advantage of a virtue of the aforementioned phase-based frequency-masking beamformer \cite{rascon2021corpus}: it not only is able to produce a preliminary estimation of the steered sound source, but also of the cumulative environmental interference \cite{rascon2021corpus}. This two-output version of the beamformer is herein referred to as \textit{beamformphasemix}, while the original one-output version is \textit{beamformphase}. With the output of \textit{beamformphasemix}, a new Demucs-based model (\textit{demucsmix}) is here proposed that employs both outputs of the beamformer, as presented in Figure \ref{fig:demucslocatemix}.

In Table \ref{tab:demucsmixresults}, the speech enhancement performance of \textit{demucs} and \textit{demucsmix} is shown in terms of PESQ \cite{recommendation2001perceptual} and STOI \cite{taal2010short} (where a higher value is better), as well as their memory usage (in megabytes).

\begin{table}[H]
	\centering
	\begin{tabular}{l|ll}
		Model & \textit{demucs}  & \textit{demucsmix} \\
		\hline
		PESQ    & 1.7599   & \textbf{1.9234}  \\
		STOI    & 0.8768   & \textbf{0.8931}  \\
		Memory (MB)    & \textbf{134.1}  & 402.5  \\
	\end{tabular}
	\caption{Comparison of Demucs models.}\label{tab:demucsmixresults}%
\end{table}

As it can be seen, \textit{demucsmix} outperforms \textit{demucs}, but it is at the cost of memory usage. To this effect, both paradigms (\textit{beamformphase} with \textit{demucs}, and \textit{beamformphasemix} with \textit{demucsmix}) are provided as part of the list of agents that conform MASA, so that the user can decide which paradigm is best for their case scenario.

It is worth pointing out that it may seem unnecessary that the model should carry out the noise decoding stage that estimates the cumulative environmental interference ($\hat{N}$ in Figure \ref{fig:demucslocatemix}), since it surely is contributing to its increase in memory usage. However, other architectures were also tested that did not do this, and their performance did not improve upon the original \textit{demucs} model. Once the model was given the task to also estimate the interference, the performance showed the improvements presented in Table \ref{tab:demucsmixresults}. The reasoning behind this phenomenon may be because the sequence modeling stage of the model requires to identify which part of the input signal is the target speech, and which is not. And to appropriately sequence-model them both, the model requires to train both complete encode-decode pipelines.

\subsection{Online Speech Quality Assessment}
\label{sec:onlinesqa}
\textbf{Agent name:} \textit{onlinesqa}.

In \cite{rascon2025direction}, the speech quality is measured by using the Squim model \cite{kumar2023torchaudio} which does not require a reference recording to provide a quality estimation (aka. a non-intrusive quality estimation). It was shown to be able to provide a quality assessment ($Q$) by estimating the signal-to-distortion ratio (SDR) of the signal in real-time, but there are two issues with the original implementation which are explored and circumvented here:

\begin{itemize}
	\item The Demucs model was originally run with an input window length ($t_i$) of $0.064$ s. However, the impact of this length was not explored, which is of interest given that it has been shown \cite{rascon2023characterization} that longer input windows provide higher quality results. 
	\item The Squim model provides very inconsistent results, with a high amount of variance, even within a single static recording. The impact of its capture window length ($t_w$) of $3.0$ s. was not explored.
\end{itemize}

In all of the following evaluations, a recording of the AIRA corpus \cite{rascon2018acoustic} was used, where the target source is positioned near $0^\circ$ and an interference is positioned near $90^\circ$. The speech sources were positioned around an equilateral-triangular microphone array, each microphone pair spaced $0.18$ m. apart. The recording has a length of $30$ s., but it was repeated $4$ times to provide ample time to evaluate the consistency of the optimization process. This recording was sampled at $48$ kHz, and was fed to the beamformer with a $1024$-sample window. It then was re-sampled at $16$ kHz in real-time (since the Demucs model and the Squim model were trained at this sample rate), and fed to the rest of the modules with the window values explored in this section. It is also important to mention that the Silero-VAD technique \cite{SileroVAD} is employed for voice activity detection, since Squim assumes that its input is solely speech.

\subsubsection{Input window length ($t_i$)}

Using the reference recording provided along with the previously described recording in the AIRA corpus, the SDR was calculated from the output of the Demucs model using different values of $t_i$. The estimated DOA ($\theta_{est}$) was set at $15^\circ$, which is in a non-optimal location, so that the quality ceiling isn't reached. The results are shown in Table \ref{tab:ti}.

\begin{table}[H]
	\centering
	\begin{tabular}{ ll }
		$t_i$ (s) & SDR (dB) \\
		\hline
		$0.064$       & $-14.29$ \\
		$0.128$       & $-14.14$ \\
		$0.256$       & $-13.80$ \\
		$\mathbf{0.512}$       & $\mathbf{-13.68}$ \\
		$1.024$       & $-14.37$ \\
	\end{tabular}
	\caption{SDR using different $t_i$ values.}\label{tab:ti}
\end{table}

As it can be seen, a $t_i$ value of $0.512$ s. provides the best SDR results. This was confirmed from subjective listening sessions, where it seemed that shorter $t_i$ values resulted in more discontinuities between $t_i$ windows. Longer $t_i$ values resulted in the Demucs model not responding fast enough given the larger amount of data that it was fed. This is an important improvement from the original $t_i$ value of $0.064$ s.

\subsubsection{Capture window length ($t_w$)}

Using the previously described recording, $\theta_{est}$ was set at $0^\circ$ (to measure the highest possible SDR), and the standard deviation of the SDR estimated by the Squim model was calculated using different values of $t_w$. The value of $t_h$ was set at $0.5$ s. to provide a stable comparison. The results are shown in Table \ref{tab:tw}.

\begin{table}[H]
	\centering
	\begin{tabular}{ ll }
		$t_w$ (s) & SDR variance (dB) \\
		\hline
		$0.5$       & $1.67$ \\
		$1.0$       & $1.01$ \\
		$2.0$       & $1.19$ \\
		$\mathbf{3.0}$       & $\mathbf{0.89}$ \\
		$4.0$       & $1.11$ \\
		$5.0$       & $1.10$ \\
	\end{tabular}
	\caption{SDR variance using different $t_w$ values.}\label{tab:tw}
\end{table}

As it can be seen, a $t_w$ value of $3.0$ s provides the least amount of SDR variation. Shorter $t_w$ values result in more variance given the small amount of data they provide, while larger $t_w$ values tend to follow the energy dynamics of the recording.

\subsection{Location Optimizer}

\textbf{Agent name:} \textit{doaoptimizer}.

The work in \cite{rascon2025direction} presents a DOA correction scheme by maximizing the speech quality as assessed by the \textit{onlinesqa} agent. It worked well in specific scenarios but, as with the \textit{onlinesqa} agent, there are also several issues with its original implementation which are explored and circumvented here:

\begin{itemize}
	\item The Adam-based optimization process establishes the speed of its adaptation with the learning rate parameter ($\eta$), the value of which was extensively explored originally. However, its value is highly dependent on the amount of time it takes to receive a new quality metric ($t_h$). Thus, the impact of different combinations of these two parameters are of interest to be explored.
	\item There is only one possible solution to the optimization task (the highest quality is at the correct DOA) and the solution space is very close to convex in the vicinity of the correct DOA. Thus, it is of interest to `reset' the optimization process to a known DOA area of high speech quality if it is failing to improve upon recent DOA history.
	\item The \textit{soundloc} agent continuously provides its own estimated DOA ($\theta_{est}$), not just at the beginning of the optimization process. This value needs to be merged to the corrected DOA result ($\theta_{corr}$) in a manner where an appropriate preference is given to either values.
\end{itemize}

\subsubsection{Time hop ($t_h$) and Learning rate ($\eta$)}
\label{sec:etath}

To bound the possible combinations of $t_h$ and $\eta$ values, first, a similar evaluation to the one in the Section \ref{sec:onlinesqa} was carried out to find the $t_h$ values that provide the least amount of SDR variance. The results are shown in Table \ref{tab:th}.

\begin{table}[H]
	\centering
	\begin{tabular}{ ll }
		$t_h$ (s) & SDR variance (dB) \\
		\hline
		$0.10$       & $1.67$ \\
		$0.15$       & $1.70$ \\
		$0.20$       & $1.69$ \\
		$0.25$       & $1.25$ \\
		$0.50$       & $1.21$ \\
		$\mathbf{1.0}$       & $\mathbf{0.81}$ \\
		$\mathbf{2.0}$       & $\mathbf{0.56}$ \\
		$3.0$       & $1.74$ \\
	\end{tabular}
	\caption{SDR variance using different $t_h$ values.} \label{tab:th}
\end{table}

As it can be seen, $t_h$ values between $1.0$ s. and $2.0$ s. provide the least amount of SDR variance. With this information, another evaluation was carried out to find the best combination of $t_h$ and $\eta$ values. To do this, for every combination, $10$ runs were carried out (obtaining the value of $\theta_{corr}$ that maximizes $Q$), with varying amounts of location errors ($\theta_{est}$). A `good run' is defined as one in which the root-mean-square of the segment of the last $3/4$ of its $\theta_{corr}$ values is in the $\pm 5^\circ$ range. This segment was chosen since the first $1/4$ of the run is where the optimization process is expected to be still converging into the correct DOA, and should not affect the result of establishing it as a `good run'. Since there was a considerable amount of evaluated combinations ($t_h = [1.0,1.5,2.0]$, $\eta = [0.2,0.25,0.3,0.35,0.4,0.5,0.6]$, $\theta_{est} = [1,5,10,15,20,25]$), only the top $10$ combinations (with the highest amount of average `good runs' throughout all $\theta_{est}$ values) are reported in Table \ref{tab:etath}.

\begin{table}[H]
	\centering
	\begin{tabular}{ ll|l }
		$t_h$ (s) & $\eta$ & \# good runs \\
		\hline
		$\mathbf{1.5}$ & $\mathbf{0.30}$ & $\mathbf{4.50}$ \\
		$2.0$ & $0.20$ & $4.17$ \\
		$1.5$ & $0.20$ & $4.17$ \\
		$1.0$ & $0.20$ & $3.67$ \\
		$2.0$ & $0.25$ & $3.50$ \\
		$1.0$ & $0.25$ & $3.33$ \\
		$2.0$ & $0.30$ & $3.17$ \\
		$1.5$ & $0.25$ & $2.83$ \\
		$0.5$ & $0.20$ & $2.83$ \\
		$1.5$ & $0.35$ & $2.67$ \\
	\end{tabular}
	\caption{Best combinations of $t_h$ and $\eta$.}\label{tab:etath}
\end{table}

As it can be seen, the best combination is $t_h$ at $1.5$ s. and $\eta$ at $0.30$. It is worth noting, however, that there is a tendency of low $\eta$ values in this list (there is only one instance of $0.35$; the rest are below that value). Additionally, the most common $t_h$ values are $1.5$ s. and $2.0$ s., with $1.5$ s. being the most common of them both. With both of these insights in conjunction, it can be concluded that $t_h$ values between $1.5$ s. and $2.0$ s. and $\eta$ values between $0.2$ and $0.3$ should provide a higher rate of success.

However, it is also worth noting that these results do vary depending on the value of $\theta_{est}$ as well. In Table  \ref{tab:etathrange}, the top $3$ combinations in three different $\theta_{est}$ ranges are shown.

\begin{table}[H]
	\centering
	\begin{tabular}{ l|lll }
		$\theta_{est}$ range & $t_h$ (s) & $\eta$ & \# good runs \\
		\hline
		$[1^\circ,5^\circ]$   &   $2.0$  & $0.20$ & $8.0$ \\
		$[1^\circ,5^\circ]$   &   $1.0$  & $0.20$ & $7.0$ \\
		$[1^\circ,5^\circ]$   &   $1.5$  & $0.20$ & $7.0$ \\
		$[10^\circ,15^\circ]$ &   $1.5$  & $0.30$ & $6.0$ \\
		$[10^\circ,15^\circ]$ &   $1.5$  & $0.20$ & $5.0$ \\
		$[10^\circ,15^\circ]$ &   $1.5$  & $0.35$ & $5.0$ \\
		$[20^\circ,25^\circ]$ &   $0.5$  & $0.25$ & $2.5$ \\
		$[20^\circ,25^\circ]$ &   $1.0$  & $0.25$ & $2.0$ \\
		$[20^\circ,25^\circ]$ &   $1.5$  & $0.30$ & $1.5$ \\
	\end{tabular}
	\caption{Best combinations of $t_h$ and $\eta$, by $\theta_{est}$ range.}	\label{tab:etathrange}
\end{table}

As it can be seen, although the aforementioned $t_h$ and $\eta$ value ranges still apply, there is a difference of effectiveness depending on the location error ($\theta_{est}$). Thus, an additional approach is required to make the DOA correction behavior more consistent. A new optimization mechanism can help with this, and is explored in the following section.

\subsubsection{New Optimization Mechanism with DOA Merger} 
\label{sec:optimizationmechanism}

The optimization technique employed in the original DOA correction scheme \cite{rascon2025direction} used a variation of the Adam technique \cite{kingma2014adam}. The bias correction was removed since it was shown to hinder the initial adaptation of the optimization process, making it `get stuck' prematurely in a non-optimal starting $\theta_{corr}$ value. However, even with this modification, a very low $\eta$ value of $0.1$ was ultimately chosen to avoid correction overshooting the correct DOA too much. Although this value provided consistent results, it was chosen using runs with only a $\theta_{est}$ value of $15^\circ$. And, as it was seen previously, higher $\eta$ values are required to obtain a generalized good performance with varying values of $\theta_{est}$. Thus, a new optimization mechanism is required to work with this new recommended $\eta$ values, and reduce overshooting.

The new optimization mechanism is such that, at any given moment, the correction process can recognize that it has overshot the correct DOA by remembering the $\theta_{corr}$ that has provided the highest $Q$ up until to that point ($Q_{best}$), and `reset' itself if necessary. Thus, if none of the recent $\theta_{corr}$ values have not provided a higher $Q$ than $Q_{best}$, the new optimization mechanism reverts back to the $\theta_{corr}$ value that provided it ($\theta_{best}$) and re-starts the optimization process from there. This new mechanism is detailed in Algorithm \ref{alg:newadam}, where all variables are initialized at $0$, unless stated otherwise.

\begin{algorithm}[H]
	\begin{algorithmic}[1]
		\Require $t_h$: time hop; $\eta$: learning rate
		\Require $\beta_m$: momentum factor ($0.9$); $\beta_v$: variance factor ($0.999$)
		\State Initialize: $\theta_{hist} \gets [0,0,\dots,0]$; $i_{ncm} \gets 10$; $\theta_{best} \gets \mathit{None}$; $Q_{best} \gets Q_{MAX}$
		\State Initialize: $\theta_c \gets \theta_{est}$ \Comment{from \textit{soundloc} agent}
		\Loop
		\State shift $\theta_{hist}$ one position
		\State $\theta_{hist}[0] \gets \theta_c$
		\State $i_{calc} \gets i_{calc}+1$
		\If{$i_{calc} \geq \mathit{length}(\theta_{hist})$}
		\If{$Q_c < Q_{best}$}
		\State $Q_{best} \gets Q_c$
		\State $\theta_{best} \gets \theta_c$
		\Else
		\State $i_{nocorr} \gets i_{nocorr}+1$ 
		\If{$i_{nocorr} \geq i_{ncm}$}
		\State $\theta_c \gets \theta_{best}$ \Comment{reset to best DOA}
		\State $\theta_p \gets 0$
		\State $\theta_{hist} \gets [0,0,\dots,0]$
		\State $\theta_{best} \gets \mathit{None}$
		\State $Q_p \gets 0$; $Q_c \gets 0$
		\State $i_{nocorr} \gets 0$; $i_{calc} \gets 0$
		\EndIf
		\EndIf
		\EndIf
		\State $\theta_c \gets merge\_theta(\theta_c, \theta_{est})$ \Comment{from \textit{soundloc} agent}
		\State $\theta_c \to \theta_{corr}$ \Comment{to speech enhancement module}
		\State wait $t_h$ seconds
		\State $Q_p \gets Q_c$
		\State $Q_c \gets Q$ \Comment{from \textit{onlinesqa} agent}
		\State $Q_c \gets quality\_norm(Q_c)$
		\State $\nabla Q \gets \frac{Q_c-Q_p}{\theta_c-\theta_p + \epsilon}$ 
		\State $\nabla m \gets \beta_m \nabla m + (1 - \beta_m)\nabla Q$ 
		\State $\nabla v \gets \beta_v \nabla v + (1 - \beta_v)\nabla Q^2$
		\State $\theta_p \gets \theta_c$
		\State $\theta_c \gets \theta_c - \eta \frac{\nabla m}{\sqrt{\nabla v} + \epsilon}$
		\EndLoop
		\caption{New optimization mechanism.}\label{alg:newadam}
	\end{algorithmic}
\end{algorithm}

There are two functions used in Algorithm \ref{alg:newadam}. The $quality\_norm$ function normalizes the value of a quality value ($Q$) such that its value is aimed to be minimized (as the Adam implementation expects it to be). This is carried out by subtracting $Q$ from its maximum value ($Q_{MAX}$): $quality\_norm(Q):= Q_{MAX}-Q$. In the case of the SDR metric, $Q_{MAX}=100$.

As for the $merge\_theta$ function, it aims to merge the DOA estimation that is continuously provided by the \textit{soundloc} agent ($\theta_{est}$) and the corrected DOA ($\theta_{corr}$). It does so by implementing the concept of `boredom' and `excitement', based on the variability of $\theta_{est}$ and $\theta_{corr}$. Basically, the weighted standard deviation of the linearly detrended $\theta_{est}$ history ($\sigma_{est_{hist}}$) establishes if there has been a recent important change in the output of the \textit{soundloc} agent. If not, the system is deemed `bored', and continues with the normal execution of Algorithm \ref{alg:newadam}, only outputting $\theta_{corr}$. If there has been an important recent change in $\sigma_{est_{hist}}$, the system is deemed `excited' and, based on the ratio between the recent variability of $\theta_{est}$ and $\theta_{corr}$, a new $\theta_{corr}$ is calculated, such that the $\theta$ with more variability is given more preference. Algorithm \ref{alg:mergedoa} details this process.

\begin{algorithm}[H]
	\begin{algorithmic}[1]
		\Require $\theta_{hist}$: history of $\theta_c$ from Alg. \ref{alg:newadam}
		\Require $\theta_{est_{hist}}$: history of $\theta_{est}$, detrended linearly \Comment{from \textit{soundloc} agent}
		\Require $i_{bored}$: current number of `bored' windows
		\Require $N_{bored}$: maximum number of `bored' windows
		\Require $\sigma_{bored}$: variance threshold to establish `boredom'
		\State $N_{est_{hist}} \gets \mathit{length}(\theta_{est_{hist}})$
		\State $w_{est_{hist}} = \frac{1}{1+e^{-\left(\left[1...N_{est_{hist}} \right]-\frac{N_{est_{hist}}}{2}\right)}}$
		\State $\overline{\theta_{est_{hist}}} \gets \frac{\sum_i \theta_{est_{hist}}[i]}{N_{est_{hist}}}$
		\State $\sigma_{est_{hist}} \gets \sqrt{\frac{\sum_i (\theta_{est_{hist}}[i] - \overline{\theta_{est_{hist}}})*(\theta_{est_{hist}}[i]w_{est_{hist}}[i] - \overline{\theta_{est_{hist}}})}{N_{est_{hist}}-1}}$
		\If{$i_{bored} \geq N_{bored}$}
		\If{$\sigma_{est_{hist}} > \sigma_{bored}$}
		\State $i_{bored} \gets 0$ \Comment{Excited, start using $\theta_{est}$}
		\State suspend Algorithm \ref{alg:newadam}
		\Else
		\State restart Algorithm \ref{alg:newadam} \Comment{Bored, keep using $\theta_{corr}$}
		\EndIf
		\State $\theta_{corr} \gets \theta_{hist}[last]$
		\Else
		\State $N_{hist} \gets \mathit{length}(\theta_{hist})$
		\State $\overline{\theta_{hist}} \gets \frac{\sum_i \theta_{hist}[i]}{N_{hist}}$
		\State $\sigma_{hist} \gets \sqrt{\frac{\sum_i (\theta_{hist}[i] - \overline{\theta_{hist}})^2}{N_{hist}-1}}$
		\State $w_{est} \gets atan \left(\frac{\sigma_{hist}}{\sigma_{est_{hist}}} \right)$
		\State $\theta_{corr} \gets (w_{est}*\overline{\theta_{hist}}) + ((1-w_{est})*\theta_{hist}[last])$
		\If{$\sigma_{hist} > \sigma_{bored}$}
		\State $i_{bored} \gets 0$ \Comment{Keep being excited}
		\Else
		\State $i_{bored} \gets i_{bored} + 1$  \Comment{Starting to get bored}
		\EndIf
		\EndIf
		\State return $\theta_{corr}$
		\caption{$merge\_theta$ function.}\label{alg:mergedoa}
	\end{algorithmic}
\end{algorithm}

To evaluate the impact of this new optimization mechanism, $10$ runs were carried out with $\eta$ set at $0.3$ and $t_h$ set at $1.5$ s. (as recommended by the results presented in Section \ref{sec:etath}), with different values of $\theta_{est}$ ($[1,5,10,15,20,25]$). Each of those sets of runs were ran with both the original optimization mechanism, as well as with the proposed new optimization mechanism. The average RMS ($\theta_{error}$), their standard deviation ($\theta_{var}$), as well as the number of good runs per each $[\eta,t_h,\theta_{est}]$ combination are presented in Table \ref{tab:optresults}.

\begin{table}[H]
	\centering
	\begin{tabular}{ l|ll|ll|ll }
		& \multicolumn{2}{c}{$\theta_{error}$} & \multicolumn{2}{c}{$\theta_{var}$} & \multicolumn{2}{c}{\# good runs} \\
		\hline
		$\theta_{est}$ & orig.  & new    & orig.  & new    & orig. & new \\
		\hline
		$1^\circ$        & $6.72$ & $\mathbf{4.96}$ & $3.02$ & $\mathbf{2.20}$ & $5$   & $\mathbf{6}$\\
		$5^\circ$        & $4.85$ & $\mathbf{3.77}$ & $2.51$ & $\mathbf{0.54}$ & $7$   & $\mathbf{10}$\\
		$10^\circ$       & $4.93$ & $\mathbf{4.02}$ & $2.41$ & $\mathbf{0.67}$ & $6$   & $\mathbf{9}$\\
		$15^\circ$       & $6.19$ & $\mathbf{5.99}$ & $3.63$ & $\mathbf{2.67}$ & $4$   & $\mathbf{5}$\\
		$20^\circ$       & $9.28$ & $\mathbf{7.03}$ & $6.27$ & $\mathbf{3.65}$ & $2$   & $\mathbf{4}$\\
		$25^\circ$       & $\mathbf{11.31}$ & $12.82$ & $\mathbf{7.33}$ & $7.34$ & $0$   & $0$\\
	\end{tabular}
	\caption{New optimization mechanism compared to the original.}\label{tab:optresults}
\end{table}

As it can be seen, in all but the case of $\theta_{est}=25^\circ$, the new optimization mechanism provides several benefits over the original mechanism, as it:

\begin{itemize}
	\item Reduces the error throughout the run (less $\theta_{error}$).
	\item Provides considerably more consistent results (less $\theta_{var}$).
	\item Increases the rate of success (more `good runs').
\end{itemize}

\begin{figure}[H]
	\centering
	\includegraphics[width=0.65\textwidth]{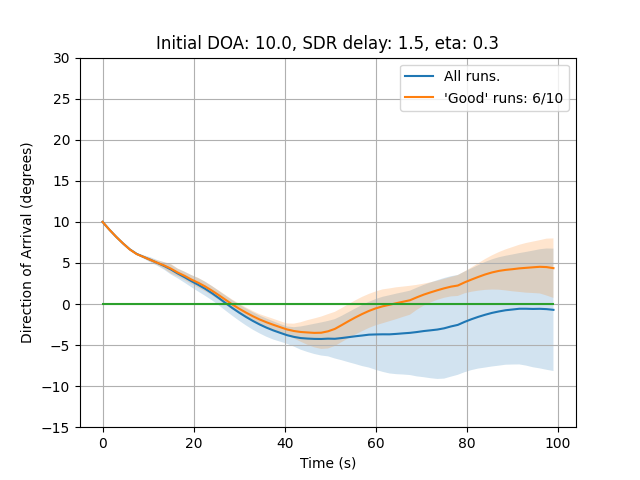}
	\caption{With the original optimization mechanism.}
	\label{fig:behaviorold}
\end{figure}

\begin{figure}[H]
	\centering
	\includegraphics[width=0.65\textwidth]{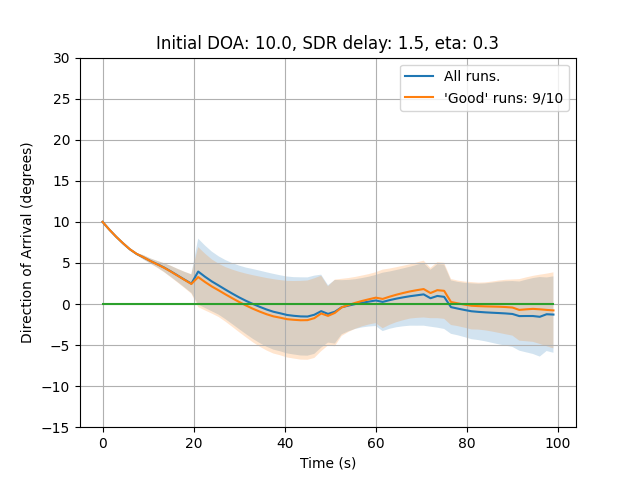}
	\caption{With the new optimization mechanism.}
	\label{fig:behaviornew}
\end{figure}

To further illustrate these benefits, Figures \ref{fig:behaviorold} and \ref{fig:behaviornew} show the original and the new optimization mechanism's behavior (respectively) with $\theta_{est}=10^\circ$. It can be seen how the variance is reduced at the end of the runs, with a more consistent stream of $\theta_{corr}$ values within the $\pm 5^\circ$ range.

As for the case of $\theta_{est}=25^\circ$, it was already established in \cite{rascon2025direction} that the original optimization mechanism is not able to recover from a such an error, so the new optimization mechanism was not expected to surpass this issue.

\subsubsection{SDR vs STOI}

\begin{figure}[H]
	\centering
	\includegraphics[width=0.7\textwidth]{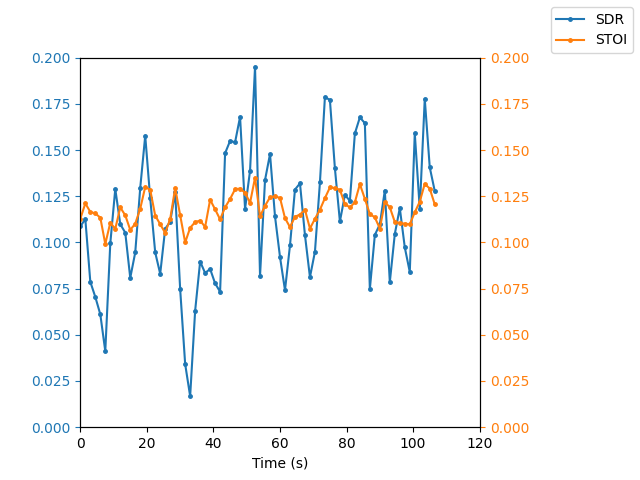}
	\caption{SDR and STOI Squim output, divided by their respective L2 norm for ease of comparison.}
	\label{fig:sdrvsstoi}
\end{figure}

The Squim model does not only provide an estimation of speech quality via the SDR metric, but also via the STOI metric \cite{taal2010short}. In Figure \ref{fig:sdrvsstoi}, a comparison of both metrics is shown through time, with the same input, with no DOA correction being carried out. Both outputs are divided by their respective L2 norm, for ease of comparison.

As it can be seen, the STOI metric provides a much less variable speech quality assessment. To test this metric with the rest of the system, however, it is necessary to also fine-tune other hyperparameters, similarly to what was carried out previously with the SDR metric. In this case, these hyperparameters include: $\eta$, the initial $\theta_{est}$, and the smoothing parameter ($\alpha$). It is important to remember that an exponential smoothing technique is used to smooth the output of the \textit{onlinesqa} agent \cite{rascon2025direction}, as shown in (\ref{eq:expsmooth}). 

\begin{equation}
\label{eq:expsmooth}
Q_k \gets \alpha Q_k + (1-\alpha)Q_{k-1}
\end{equation}

The smoothing parameter $\alpha$ was fine-tuned along with other hyperparameters in the original work \cite{rascon2025direction}, and, thus, if a new metric is to be used, this parameter needs to be fine-tuned as well.

\begin{figure}[H]
	\centering
	\includegraphics[width=0.7\textwidth]{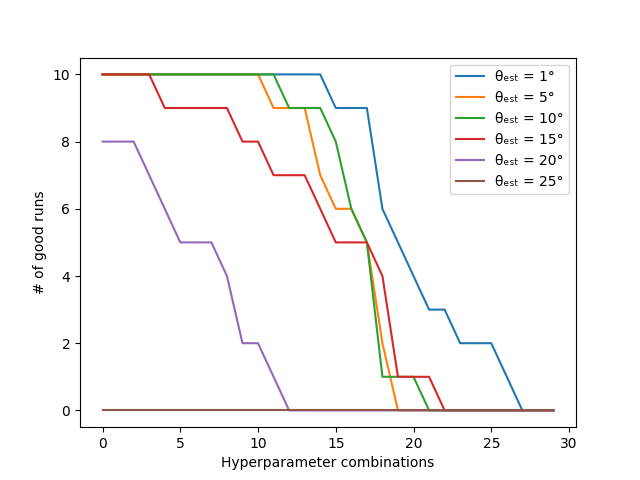}
	\caption{Number of good runs with different hyperparameter combinations, with varying values of $\theta_{est}$.}
	\label{fig:goodruns}
\end{figure}

Similarly to previous fine-tuning efforts in this work, there was a considerable amount of evaluated combinations ($\alpha = [0.40,0.50,0.60,0.70,0.80,0.90]$, $\eta = [0.2,0.3,0.4,0.5,0.6]$, $\theta_{est} = [1,5,10,15,20,25]$). Since these evaluations are already using the new optimization scheme, it is important to find a set of hyperparameters that work well in most scenarios (meaning, with varying values of $\theta_{est}$). In Figure \ref{fig:goodruns}, the number of good runs are shown that were obtained with different hyperparameter combinations, ordered in descending number of good runs.

As it can be seen, there is a substantial drop in the number of good runs depending on the value of $\theta_{est}$, except for $\theta_{est} = 25^\circ$ (which will be ignored from this point on, since the proposed system is still not able to correct such a high amount of location error). Thus, only the hyperparameter combinations that are above that drop (depending on the value of $\theta_{est}$) are here considered as `good' hyperparameter combinations. From these `good' combinations, a weighted average for $\eta$ and $\alpha$ was calculated (using the number of good runs as weight), and are shown in Table \ref{tab:goodruns}.

\begin{table}[H]
	\centering
	\begin{tabular}{ l|ll }
		$\theta_{est}$ & $\eta$ & $\alpha$ \\
		\hline
		$1.0^\circ$ & $0.2983$ & $0.6520$ \\
		$5.0^\circ$ & $0.2888$ & $0.6590$ \\
		$10.0^\circ$ & $0.3000$ & $0.6494$ \\
		$15.0^\circ$ & $0.3993$ & $0.6592$ \\
		$20.0^\circ$ & $0.5393$ & $0.6446$ \\
	\end{tabular}
	\caption{Weighted averaged $\eta$ and $\alpha$ per $\theta_{est}$.} \label{tab:goodruns}
\end{table}

As it can be seen, a value for $\alpha$ between $0.6$ and $0.7$ is used across many `good' hyperparameter combinations, across all the evaluated $\theta_{est}$ values. As for $\eta$, a value between $0.3$ and $0.4$ seems to work well with $\theta_{est} \leq 15.0^\circ$. When $\theta_{est}=20^\circ$, a higher $\eta$ value of $\sim 0.5$ is necessary to reach the correct DOA before the end of the recording. However, it is important to note that the correction mechanism does work with $\eta=0.3$ when $\theta_{est}=20^\circ$, it just takes considerably more time.

\begin{figure}[H]
	\centering
	\includegraphics[width=0.65\textwidth]{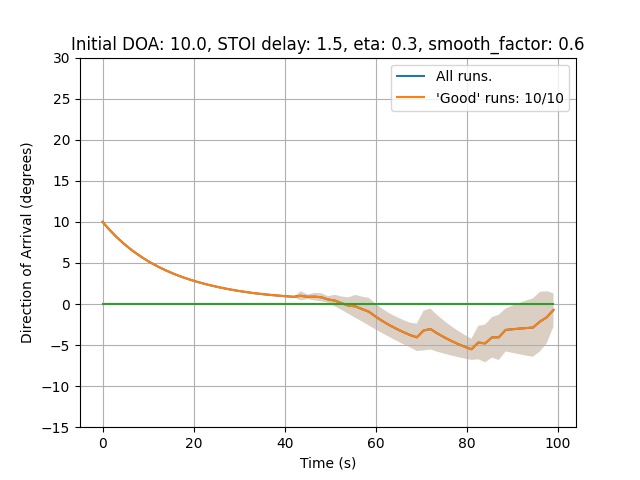}
	\caption{With the new optimization mechanism and using STOI (instead of SDR).}
	\label{fig:behaviornewstoi}
\end{figure}

To better visualize the improvement of DOA correction behavior when using STOI, in Figure \ref{fig:behaviornewstoi} this behavior is shown with the same configuration as in Figure \ref{fig:behaviornew}.

As it can be seen, the amount of good runs are increased, while the amount of variability in the $\theta_{corr}$ values is substantially reduced when using STOI compared to when using SDR. It is also important to mention that $Q_{MAX}$ in Algorithm \ref{alg:newadam} is $1$ when the STOI metric is used for quality assessment.

However, it is important to note that while STOI provides a more consistent behavior, it does seem that SDR is more responsive to changes in location. This is important in cases where the source of interest is mobile. To this effect, both options are provided in the MASA implementation, and it is left to the user to decide which quality metric is more appropriate to their application scenario.

\subsection{Frequency Selection based on Source Type}
\label{sec:freqselect}
\textbf{Agent name:} \textit{freqselect}.

Many localization techniques search for sound sources throughout the whole frequency spectrum \cite{rascon2017localization}, which may not only make them difficult to run in real-time, but may introduce errors if there are any interferences in irrelevant frequencies. Thus, other localization techniques focus their search on a given frequency range of interest via pre-filtering \cite{rascon2015lightweight} or by frequency masking the noise/interference  \cite{grondin2016noise}. However, both of these types of approaches assume that the frequency components of either the source of interest (to be focused on) or the interference (to be ignored) are known, and that are either static or change slowly through time.

The \textit{freqselect} agent carries out a dynamic frequency selection based on the type of sound source that is present in the input audio mixture. In this case, urban sounds are to be ignored, requiring a small database with the dominant frequencies of each class of urban sound. Accordingly, the \textit{soundloc} agent is modified such that it masks its inputs with the frequencies fed by the \textit{freqselect} agent, as to nullify them.

As to how the sound source is classified, a simple support-vector-machine model has been trained to classify between four urban sounds, using as its input a set of mel-frequency cepstral coefficients \cite{davis1980comparison} calculated from the last $2$ s. window captured from the reference microphone. This model was chosen only for demonstration purposes, but it is noteworthy that it is very lightweight compared to its deep-learning-based counterparts \cite{massoudi2021urban,nogueira2022sound}, which are more complex but also more robust. However, it is important to note that, if the user requires it, the chosen model can be easily replaced with these other models in the \textit{freqselect} agent.

\begin{figure}[H]
	\centering
	\includegraphics[width=0.7\textwidth]{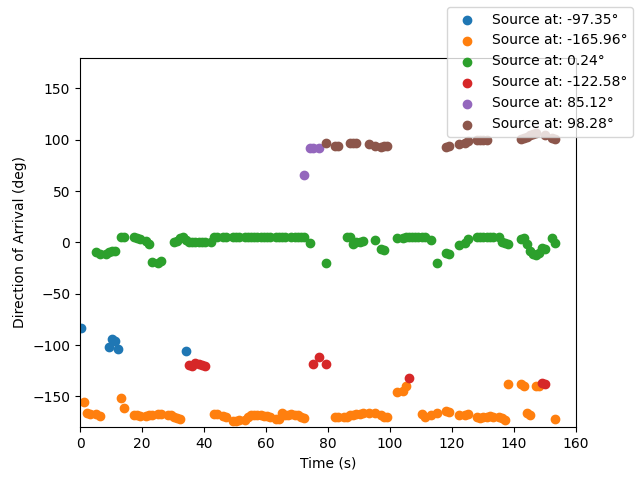}
	\caption{Output of the \textit{soundloc} agent with noise.}
	\label{fig:soundloc_withnoise}
\end{figure}

Both the frequency selection and the urban sound classification, working in conjunction, effectively implement another feedback loop in the proposed multi-agent approach. To demonstrate the effectiveness of this feedback loop, a 'car horn' type of urban sound was artificially added as an interference to the input audio mixtures. In Figure \ref{fig:soundloc_withnoise}, the output of the \textit{soundloc} agent is shown without using the feedback loop, with all DOA tracks that it estimated (represented by their average in degrees).

\begin{figure}[H]
	\centering
	\includegraphics[width=0.7\textwidth]{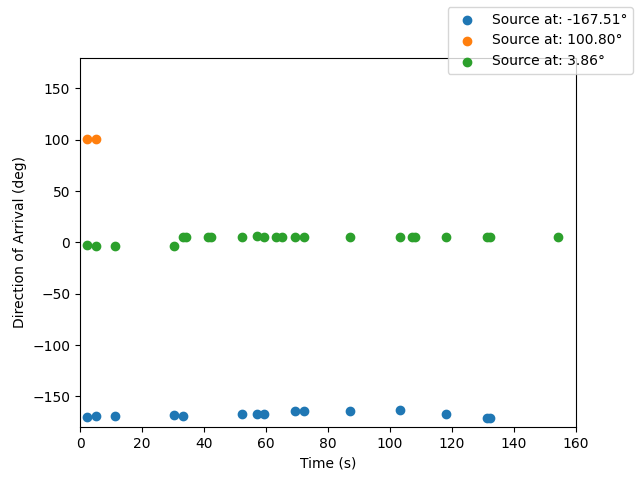}
	\caption{Output of the \textit{soundloc} agent with frequency selection.}
	\label{fig:soundloc_filtered}
\end{figure}

In Figure \ref{fig:soundloc_filtered}, the output of the \textit{soundloc} agent is shown when it masked the peak frequencies of the urban sound, which are: $999$, $1229$, $1485$, $1630$, $2005$, $2478$, $2844$, $2985$, $3257$, $3486$, $3670$, and $3976$ Hz. These were extracted from the log-frequency domain, calculated from the urban sound recording that was inserted into the input audio mixture.

As it can be seen, when the interference is inserted (Figure \ref{fig:soundloc_withnoise}) the speech source of interest at $0^\circ$ is consistently accurately located, but it does so along with a considerable amount of other sound sources. However, when the noise is masked (Figure \ref{fig:soundloc_filtered}), the presence of the other sound sources is greatly reduced. Although the amount of location estimations of the source of interest is also reduced, the reader is reminded that the new optimization mechanism of the \textit{doaoptimizer} (detailed in Section \ref{sec:optimizationmechanism}) employs a DOA merging method (as described in Algorithm \ref{alg:mergedoa}) which requires only sporadic results from the \textit{soundloc} agent ($\theta_{est}$).

It is also important to mention that the case scenario here presented can be reversed, where the \textit{freqselect} agent can select the frequencies which the \textit{soundloc} needs to focus on. To achieve this, only trivial modifications to both agents are required.

\section{Implementation}\label{sec:implementation}

A brief summary of the tools used to run the proposed multi-agent approach is presented here. For one, this is due to satisfy completeness in the description of the approach. But, additionally, to serve the user as a guide to the software requirements necessary to run MASA.

Audio acquisition and reproduction is carried out by the JACK Audio Connection Toolkit (JACK) \cite{JACKAudio}. It was chosen as it has been extensively used in multi-channel audio processing scenarios \cite{letz2005jackdmp} where low latency is essential \cite{newmarch2017jack}, which is the case for real-time auditory scene analysis. Additionally, the latest version of JACK runs very much like a MAS \cite{letz2009s}, which is compatible to the nature of the proposed approach.

It is important to note, however, that JACK ultimately forces a linear data flow between its agents, and is not developed specifically for the bi-directional flow (aka. feedback) that MASA requires. And, obviously, it does not provide any type of pipelines to communicate non-audio information between agents. Thus, another communication protocol is required to carry out the feedback of non-audio information. For these purposes, the ROS2 protocol \cite{ROS2} is used. It has been widely used for near real-time inter-agent communication \cite{macenski2022robot} in fields such as robotics and general automation \cite{reke2020self}. It has also shown viability for low-power applications \cite{maruyama2016exploring}. Finally, and more importantly, it is ideal to be used in a MAS since it has shown good performances in collaborative and intelligent automation systems \cite{erHos2019ros2}. Because of the open nature of both protocols (JACK and ROS2), as well as the adaptive nature of a MAS, it is possible to add new agents to MASA by the users, depending of their case scenario.

To this effect, JACK is applied for the `end' agents that deal directly with the acquisition and/or reproduction of audio, such as \textit{soundloc}, \textit{beamformphase}, \textit{freqselect}, and \textit{demucs}. ROS2 is used for communication between `middle' agents that deal with non-audio information, such as \textit{onlinesqa} and \textit{doaoptimizer}. As it can be deduced, several `end' agents also deal with non-audio information, such as: \textit{soundloc}, which acquires audio but provides $\theta_{est}$; \textit{beamformphase}, which acquires and reproduces audio but requires $\theta_{est}$; and \textit{freqselect}, which acquires audio but provides a frequency mask. In these cases, agents employ both the JACK and ROS2 protocols in parallel. Since JACK has very strict latency requirements, a `bridge' between both protocols (referred here as \textit{ros2jack}) was implemented. It employs a $\sim 1$ s. buffer to be used in cases of ROS2 packet loss. These losses may result in latency issues which will ultimately produce catastrophic JACK2 overruns if not for this buffer. Surprisingly, ROS2 has shown that it is able to `keep up' with JACK2's strict latency requirements by its own, evidenced by the fact that the aforementioned buffer was rarely utilized in the previous evaluations. However, since this was not a result that was being tested for, it is left for future work to analyze the effect and actual need (if any) of this buffer.

Other agents have been also implemented (and included in the public repository available at: \url{https://github.com/balkce/masa}) with the objective of facilitating the act of running all of the MASA agents, as well as to visualize and listen to its results:

\begin{itemize}
	\item \textit{doa\_plot}: real-time visualization of \textit{soundloc}'s output: $\theta_{est}$.
	\item \textit{theta\_plot}: real-time visualization of \textit{doaoptimizer}'s output $\theta_{corr}$.
	\item \textit{qual\_plot}: real-time visualization of \textit{onlinesqa}'s output: $Q$.
	\item \textit{jack\_write}: real-time audio of the speech enhanced by \textit{demucs}.
	\item \textit{masacoord}: provides the function of running all or some of the MASA agents, in a given order, with one command.
\end{itemize}

The \textit{masacoord} agent relies on the Terminator \cite{Terminator} terminal emulator to visualize all agent's real-time log information in one window. The user provides a list of agents to launch (trivially registered in a configuration file) in a given order. Then, \textit{masacoord} creates, for each of these agents, its own emulated terminal, such that it can be terminated and re-initialized easily without requiring restarting any other agent. Finally, all agents can be all terminated at once only by terminating the \textit{masacoord} agent.

\section{Results}\label{sec:results}

To evaluate the system as a whole, the same recording as the one employed in Section \ref{sec:freqselect} is used, meaning, the one in which the test recording is contaminated with a ``car horn'' type of interference.

Three different configurations of MASA are evaluated:

\begin{itemize}
	\item \textbf{Linear}: only the \textit{soundloc}, \textit{beamformphase} and \textit{demucs} agents are run, with no feedback loops. The DOA with the most confidence value from \textit{soundloc} ($\theta_{est}$) is fed directly to \textit{beamformphase}, acting as a simulated $\theta_{corr}$.
	\item \textbf{FrequencySelection}: the \textit{freqselect} agent is run along with the agents in the Linear configuration, with the classification model providing the frequencies which are to be masked by \textit{soundloc}.
	\item \textbf{DOACorrection}: in addition to all the agents in the FrequencySelection configuration, the \textit{onlinesqa} and \textit{doaoptimizer} agents are also run. A real-time correction of $\theta_{est}$ is carried out to provide $\theta_{corr}$, by optimizing the speech quality estimated by \textit{onlinesqa}, employing the new optimization mechanism of Algorithm \ref{alg:newadam} and the DOA merging technique of Algorithm \ref{alg:mergedoa}. This can be considered as the full proposed MASA approach.
\end{itemize}

In Table \ref{tab:globalresults}, the average location error as well as several speech quality metrics are provided, calculated from the last 30 s. of the recording. This segment was chosen to compare all three configurations in a steady state.

\begin{table}[H]
	\centering
	\begin{tabular}{ l|llll }
		Config. & avg. error ($^\circ$) & STOI & PESQ & SDR (dB) \\
		\hline
		Linear             & $90.1557$ & $0.7173$ & $1.1617$ & $-0.1961$ \\
		FrequencySelection & $5.4038$  & $0.8726$ & $1.3325$ & $12.5726$ \\
		DOACorrection      & $\mathbf{2.0378}$  & $\mathbf{0.8774}$ & $\mathbf{1.5122}$ & $\mathbf{16.7946}$ \\
	\end{tabular}
	\caption{Comparison between configurations.}\label{tab:globalresults}
\end{table}

As it can be seen, the amount of location error is greatly reduced once the frequency selection feedback loop is employed, which results in a greatly increased speech quality across all metrics. In addition, once the quality feedback loop is employed in addition to the classification loop, the location error is further reduced, accompanied with an even greater speech quality. This clearly shows how the use of several feedback loops benefits the location estimation and speech quality, in conjunction.

\subsection{Discussion}

The results presented here are the more impressive considering that the hardware in which all of these tests were run can be considered ``moderate'' compared to current hardware trends: a 3.6 GHz 6-Core CPU (AMD Ryzen 5 3600) and a GPU with only 4 GB of VRAM (GTX 1050 Ti). As a whole, MASA did not use more than 1 GB of memory.

In addition, it is important to note that, given the parallel nature of a MAS, the response time of the slowest agent is equivalent to the response time of the whole system. In the original work \cite{rascon2025direction}, the slowest agent was \textit{onlinesqa} which is still the case. Thus, the response time of this agent, and that of the whole system, continues being only between 0.0538 s. and 0.0704 s.

Finally, it is important to note that the results obtained with the DOACorrection configuration employed the STOI metric estimated by the Squim model. The reader is reminded that the SDR metric can also be used, however, both metrics have their cons and pros. SDR is more responsive to source movements, which may lead to better performance in dynamic environments. STOI is more stable, which may lead to better performance in static environments. To this effect, both alternatives are provided as part of MASA.

\section{Conclusion}\label{sec:conclusions}

The tasks involved in auditory scene analysis (ASA) are localization, separation and classification, and they are usually carried out such that the data flow is linear. This is very sensitive to estimation errors in the tasks at the beginning of the data flow (with special focus on localization).

In this work, a new approach to carry out ASA is proposed, by structuring it as a multi-agent system (MAS). This approach formalizes a previous work where localization errors are corrected in real-time by maximizing speech quality through an Adam-based optimizer that feeds back the corrected location to the separation model. This formalization is presented here as a multi-agent auditory scene analysis (MASA) framework, where additional feedback loops can be implemented, providing general robustness to local errors as an emerging behavior. This framework is publicly available at \url{https://github.com/balkce/masa}.

All currently implemented agents were described, along with new improvements presented as part of this work:

\begin{itemize}
	\item Localization (\textit{soundloc}): uses a more sophisticated clustering technique.
	\item Speech enhancement (\textit{demucs}): its hyperparameters were fine-tuned, and now uses both estimations of the source of interest and of the cumulative environmental interference.
	\item Online speech quality assessment (\textit{onlinesqa}): its hyperparameters were fine-tuned, and can now use both the SDR and STOI metrics, that are able to work well in dynamic and static environments, respectively.
	\item Location optimization (\textit{doaoptimizer}): its hyperparameters were fine-tuned, and a new optimization mechanism is employed that remembers the location with the best quality. It also now merges the corrected location with the one provided by \textit{soundloc}, preferring the one with more variance to react quickly to changes in the environment.
	\item Frequency selection by source type (\textit{freqselect}): a new feedback loop that selects frequencies to ignore during localization, depending on the type of urban sound interference that is in the environment.
\end{itemize}

Results show a considerable reduction in location errors when using the frequency selection agent, which effectively carries out a feedback loop from the classification task to the localization task. This in turn results in a considerable increase in speech quality. By employing the location optimization agent, the location error is further reduced, resulting in a further increase in speech quality. This tendency explicitly demonstrates that the feedback-based nature of MASA delivers robustness against local errors.

In addition to this benefit, it is important to note that all the agents employ relatively simple techniques (compared to their state-of-the-art counterparts) to carry out their tasks. This results in the system overall, in addition to being robust, baring a small computational footprint, along with having very low response times.

The proposed framework used open source, freely available software libraries for audio acquisition (JACK) and inter-agent communication (ROS2), enabling further additions of new agents by the user (for their specific case scenario) as well as by the ASA community at large.

As for future work, the buffer used on the JACK-ROS2 bridge is in need of characterization. In addition, there is still another feedback loop to be explored: feeding the selected frequencies to the speech enhancement agent to either further focus on the source of interest or further remove interferences. Furthermore, the current implementation of the system is focused on locating and separating speech sources, however, it is also of interest to also locate, separate, and classify other types of sound sources. Finally, the convergence time may be improved with techniques such as those of adaptive control engineering \cite{shtessel2014sliding,wang2009fast}. 

\section*{Acknowledgements}

The authors would like to acknowledge Alexandre D\'{e}fossez, Gabriel Synnaeve, and Yossi Adi, the creators of the original Demucs-Denoiser model \cite{defossez20_interspeech}, for their efforts in making their implementation publicly accessible at \url{https://github.com/facebookresearch/denoiser}.

\section*{Author Contributions}

Conceptualization; Formal analysis and investigation; Writing - original draft preparation; Funding acquisition; Resources; and Supervision: Caleb Rascon. Methodology; Writing - review and editing: Caleb Rascon, Luis Miguel Gato-D\'{i}az, Eduardo Garc\'{i}a-Alarc\'{o}n.

\section*{Funding} This work was supported by PAPIIT-UNAM through the grant IN100624.

\bibliographystyle{elsarticle-num} 
\bibliography{bibliography}

\end{document}